\documentstyle[psfig,floats,twocolumn,prl,aps]{revtex}
\begin{document}
\title{Broadening of Andreev Bound States in d$_{x^2-y^2}$ superconductors}
\author{A. Poenicke$^1$, Yu.~S. Barash$^2$, C. Bruder$^1$, and V. Istyukov$^2$}
\address{$^1$ Institut f{\"u}r Theoretische Festk{\"o}rperphysik,
Universit{\"a}t Karlsruhe, D-76128 Karlsruhe, Germany\\
$^2$ P.N. Lebedev Physical Institute, Leninsky Prospect 53, Moscow 117924,
Russia}
\date{\today}
\maketitle
\begin{abstract}
We investigate the broadening of the bound states at an interface of
an unconventional superconductor by bulk impurity scattering.  We use
the quasiclassical theory and include impurity scattering in the Born
and in the unitarity limit. The broadening of bound states due to
unitary scatterers is shown to be substantially weaker than in the
Born limit. We study various model geometries and calculate the
temperature dependence of the Josephson critical current in the
presence of these impurity-broadened bound states.
\end{abstract}
\pacs{74.50.+r}


\section{Introduction}
After a lot of discussion about the nature of the order parameter in
the high-T$_C$ materials, a number of experiments has made it
virtually certain that YBa$_2$Cu$_3$O$_{7-x}$ (YBCO) is a
superconductor whose order parameter has a d-wave symmetry. This was
established by phase-sensitive experiments
\cite{vanharlingen93,tsuei94} that investigated SQUIDs or ring
structures containing junctions either between YBCO and ordinary
superconductors, or between different domains of YBCO.

In unconventional superconductors like those with a d-wave symmetry,
the order parameter is sensitive to scattering from non-magnetic
impurities and surface roughness. The order parameter also has a
non-trivial structure close to surfaces and interfaces. Surfaces and
interfaces can be pair-breaking, i.e., the order parameter is
suppressed on a length scale given by the coherence length. Andreev
reflection processes of quasi-particle excitations from the spatial
profile of the order parameter, combined with conventional (and/or
Andreev) reflection from the surface (interface) can result under
certain conditions in surface (interface) bound states. In particular,
it has been shown\cite{hu} that an interface separating two d-wave
superconductors with different signs of the order parameter (in a
given $\bbox{k}$-direction) will {\it always} support a bound state at
the Fermi energy. A mathematically equivalent situation arises if we
consider a specularly reflecting wall and the sign of the order
parameter is different for incoming and outgoing quasiparticles, i.e.,
there will also be a bound state.

These bound states lead to a qualitative change of the Josephson
current through a tunnel junction as has been recently found
in\cite{barash96,tanaka96}. It was shown that the temperature
dependence of the critical current is dramatically different from the
standard Ambegaokar-Baratoff prediction. The critical current may
increase substantially at low temperatures\cite{barash96} or even
change sign, i.e., the junction may change its nature from an ordinary
junction to a $\pi$-junction characterized by a current-phase relation
$I=I_C\sin(\phi+\pi)=-I_C\sin(\phi)$. The authors of Ref.
\onlinecite{barash96} also considered the influence of surface
roughness on this phenomenon and showed that it gets weaker because of
a broadening of the bound states. In the present paper, we want to
discuss how scattering from bulk impurities changes the bound states
and, as a consequence, the Josephson critical current.

We use the quasiclassical formalism of superconductivity to obtain our
results. In Section \ref{sec:model} we briefly describe the
formalism. We include bulk impurities in the standard way by using an
impurity self-energy. There are different models for impurity
scattering, and we concentrate on two limits, viz., the limits of Born
scattering (weak scattering, scattering phase shift $\delta_0 \ll 1$)
and unitary scattering (strong scattering, $\delta_0 \to \pi/2$).

In Section \ref{sec:results} we calculate the angle-resolved density
of states and the Josephson critical current in the presence of
impurity-broadened bound states. We find by numerical calculation that
bulk impurities will cut off the zero-temperature divergencies in the
critical current predicted for clean systems. We also develop an
analytical understanding of these results and give analytical
expressions for small scattering rates. Surprisingly, we find that the
bound states are more sensitive to Born scatterers than to unitary
scatterers at a given scattering rate.

\section{Formalism and model assumptions}
\label{sec:model}
\begin{figure}[htbp]
\begin{center}
\leavevmode
\psfig{figure=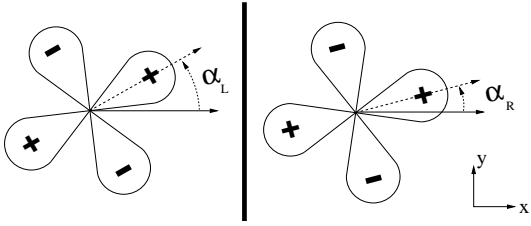,width=7cm}
\narrowtext
\vspace{0.3cm}
\caption{Geometry of the model system: a tunnel junction between two
misoriented d$_{x^2-y^2}$ superconductors}
\label{fig:geometry}
\end{center}
\end{figure}

In this paper, we consider the system shown in
Fig.~\ref{fig:geometry}, consisting of a junction between two
d$_{x^2-y^2}$ superconductors. The order parameter on side $i$,
$i=L,R$ is rotated by $\alpha_i$ with respect to the surface
normal. The junction is assumed to be weakly transparent and will be
modeled as a tunnel junction. We assume that the superconductors
contain a small concentration of impurities, and we would like to
study the influence of these impurities on the quasiparticle bound
states formed at the junction\cite{hu,barash96}.

In unconventional superconductors, the order parameter is spatially
inhomogeneous close to obstacles like surfaces or interfaces. The
quasiclassical formalism of
superconductivity\cite{eilenberger,larkin,rainersauls} is ideally
suited to calculating the structure of the order parameter in a
self-consistent way. Its central object, the energy-integrated Green's
matrix function $\hat{g}(\bbox{R},\bbox{p}_{\text{F}};\varepsilon_n)$
fulfills the Eilenberger equation (we have put $\hbar=1$)
\begin{eqnarray}
\label{eilenbergereq}
[i\varepsilon_n\hat{\tau}_3-\hat{\Delta}(\bbox{R},\bbox{p}_{\text{F}})-&&
\hat{\Sigma}(\bbox{R},\varepsilon_n),\hat{g}
(\bbox{p}_{\text{F}},\varepsilon_n)]
\nonumber\\
&&+i\bbox{v}_{\text{F}}\bbox{\nabla}_{\bbox{\text{R}}}
\hat{g}(\bbox{p}_{\text{F}},\varepsilon_n)=0\; .
\end{eqnarray}

Here $\varepsilon_n=\pi T(2n+1)$ are the Matsubara frequencies,
$\hat{\Sigma}(\bbox{R},\varepsilon_n)$
is the impurity self-energy and $\hat{\tau}_i$ are the Pauli matrices.

The Green's function has to obey the normalization condition

\begin{equation}
\label{eq:normalization}
\hat{g}^2(\bbox{p}_{\text{F}},\varepsilon_n)=-\pi^2\hat{1}\; ,
\end{equation}
and $\hat{\Delta}(\bbox{p}_{\text{F}})$ and
$\hat{g}(\bbox{R},\bbox{p}_{\text{F}};\varepsilon_n)$ are related by
the self-consistency relation (we have put $k_B=1$)

\begin{equation}
\label{eq:self-consistency}
\Delta(\bbox{R},\bbox{p}_{\text{F}})=T\sum_{\varepsilon_n}
\langle V(\bbox{p}_{\text{F}},\bbox{p}_{\text{F}}')
f(\bbox{R},\bbox{p}_{\text{F}}';\varepsilon_n)\rangle_{\bbox{p}_{\text{F}}'}
\; .
\end{equation}
Here $\langle\dots\rangle$ denotes averaging over the Fermi surface,
which we assume to be cylindrical.
The model pairing interaction is defined by
\begin{equation}
\label{eq:Wechselwirkung}
V(\bbox{p}_{\text{F}},\bbox{p}_{\text{F}}')=
\lambda\cos(2\phi-2\alpha_i)\cos(2\phi'-2\alpha_i)\; ,
\end{equation}
where $i=L,R$, and $\phi$ is the azimuthal angle between $\bbox{p}_{\text{F}}$
and the surface normal $\hat{\bbox{n}}$.

The self-energy $\hat{\Sigma}$ describes impurity scattering, and we
consider two models, viz., weak scattering (scattering phase shift
$\delta_0 \ll 1$), which will be treated in the Born approximation,
and unitary scattering ($\delta_0 \to \pi/2$). For an isotropic
point-like impurity potential, the off-diagonal components of the
self-energy vanish for order parameters transforming according to
non-trivial representations of the point symmetry group of the
crystal. This is the case for the $d$-wave order parameter studied
here. The self-energy is then characterized by a single scalar
function. In the Born limit the self-energy is given by
\begin{equation}
\label{eq:selfenergy-Born}
\Sigma(\bbox{R};\varepsilon_n)
=\frac{\displaystyle \Gamma_b}{\displaystyle \pi}\left\langle
g(\bbox{R},\bbox{p}_{\text{F}};\varepsilon_n)
\right\rangle \enspace, \qquad \Gamma_b=\frac{\displaystyle 1}{\displaystyle
2\tau}
\; ,
\end{equation}
whereas in the unitary limit
\begin{equation}
\label{eq:Selbstenergie-Unitary}
\Sigma(\bbox{R};\varepsilon_n)
=\Gamma_u\frac{\displaystyle \pi}{\displaystyle \left\langle
g(\bbox{R},\bbox{p}_{\text{F}};\varepsilon_n)
\right\rangle} \enspace, \qquad \Gamma_u=\frac{n_i}{\pi N_0}\; .
\end{equation}
Here, $n_i$ is the concentration of impurities, and $N_0$ is the (normal)
density of states at the Fermi energy.

In the framework of the quasiclassical formalism, interfaces are
taken into account by Zaitsev's boundary conditions \cite{zaitsev}. The
properties of the barrier are characterized by the transmission
probability $D(\bbox{p}_{\text{F}})$.

\begin{figure}[htbp]
\begin{center}
\leavevmode
\psfig{figure=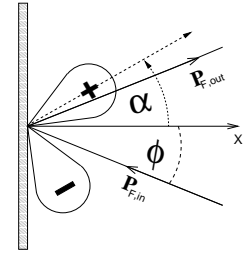,width=3cm}
\narrowtext
\vspace{0.3cm}
\caption{Geometry of incoming and outgoing quasiparticle trajectories for an
intransparent interface.}
\label{intransparent}
\end{center}
\end{figure}

In the limit of zero transparency (illustrated in
Fig.~\ref{intransparent}) the boundary conditions reduce to
$\hat{g}(\bbox{p}_{\text{F}in},0+)=\hat{g}(\bbox{p}_{\text{F}out},0+)$.
We will assume specular reflection.

We are interested in the Josephson current through the system shown in
Fig.~\ref{fig:geometry}. In first order in the transparency this can
be expressed as
\begin{eqnarray}
\label{eq:critical-current}
j_C=&&-\frac{eN_0^LT}{2\pi}\sum_{\varepsilon_n}\left\langle
v_{\text{F}_\bot}^L(\bbox{p}_{\text{F} in}^L)D(\bbox{p}_{\text{F} in}^L)
\right.\nonumber\\
&&\left[
f^{L}(\bbox{p}_{\text{F} in}^L;\varepsilon_n)
f^{+,R}(\bbox{p}_{\text{F} out}^R;\varepsilon_n)\right.\nonumber\\
&&+\left.\left.
f^{+,L}(\bbox{p}_{\text{F} in}^L;\varepsilon_n)
f^{R}(\bbox{p}_{\text{F} out}^R;\varepsilon_n)
\right]\right\rangle_{\bbox{p}_{\text{F} in}^L}\; ,
\end{eqnarray}
where the Green's functions have to be evaluated at the interface for real
order parameters, as if they would not have complex phases. In the results, we
will eliminate the transparency and express it through the normal-state
resistance of the junction which is given by (again in first order of the
transparency)
\begin{equation}
\label{eq:Normalwiderstand}
R_N^{-1}=e^2A N_0^L
\left\langle
v_{\text{F}_\bot}^L(\bbox{p}_{\text{F} in}^L)D(\bbox{p}_{\text{F} in}^L)
\right\rangle
_{\bbox{p}_{\text{F} in}^L}\; .
\end{equation}
In the calculation, we will assume the following model directional
dependence of the transparency: 
$ D(\phi)=D_0\cos(\phi)^2$.

\section{Results for Born and unitary limit}
\label{sec:results}
\begin{figure}[htbp]
\begin{center}
\leavevmode
\psfig{figure=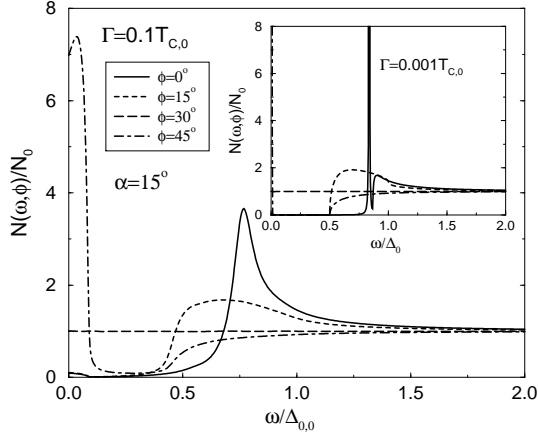,width=7cm,angle=-90}
\narrowtext
\caption{Local density of states calculated for a misorientation angle of
$\alpha=15^\circ$ at a scattering rate of $\Gamma_b=0.1T_{C,0}$
in Born approximation. The temperature is $T=0.01T_{C,0}$.
Inset: The same for a low scattering rate $\Gamma_b=0.001T_{C,0}$.}
\label{densostat}
\end{center}
\end{figure}

A surface of an unconventional superconductor may support a bound
state at the Fermi energy\cite{hu,barash96,barash97}. We reproduce
this phenomenon by a self-consistent solution of a real-time version
of Eqs.~(\ref{eilenbergereq}) - (\ref{eq:self-consistency}) taking
into account impurity scattering. Figure~\ref{densostat} shows the
angle-resolved local density of states (taken at the surface) obtained
in the Born limit for the geometry shown in
Fig.~\ref{intransparent}. The bound states present in a clean system
can be seen to be broadened by impurity scattering.

In order to estimate the dependence of the height of the zero-energy peak
in the density of states on the relaxation time in the Born limit, we
proceed analogously to the analytical consideration developed
in\cite{barash972}, based on the Eilenberger equations. In particular,
the residue of the Green's function for midgap states were analytically
calculated there in the clean limit. Due to the presence of impurities the
pole of the retarded Green's function at zero energy moves into the
complex plane, although its midgap value at the surface
$g(x=0,\bbox{p}_{\text{F}};\varepsilon=0)\equiv
g(\bbox{p}_{\text{F}},0)$ is still large in the case of a sufficiently
large relaxation time. Under this condition we obtain the following
integral equation for $g(\bbox{p}_{\text{F}},0)$:
\begin{eqnarray}
&&g(\bbox{p}_{in,\text{F}},0)\times\nonumber\\
&&\int\limits_0^\infty dx\left\{\left\langle g(\bbox{p'}_{\text{F}},0)
\exp\left(-\int\limits_0^x dx'
\frac{\displaystyle2|\Delta(\bbox{p'}_{\text{F}},x')
|}{\displaystyle |v'_x|} \right)\right\rangle_{\bbox{p}_{\text{F}}'}
\right.
\nonumber\\
&&\times\left[\exp\left(-\int\limits_0^x dx'\frac{\displaystyle2|
\Delta(\bbox{p}_{in,\text{F}}, x')|}{\displaystyle |v_x|} \right) +
\right.
\nonumber\\
&&\left.
\left.
\exp\left(-\int\limits_0^x dx'\frac{\displaystyle
2|\Delta({\bbox{p}}_{out,\text{F}},
x')|}{\displaystyle |v_x|} \right)\right]\right\}
=-2\pi^2|v_x|\tau
\enspace .
\label{integequat}
\end{eqnarray}
The order parameter can be factorized in the form
$\Delta(\bbox{p}_{\text{F}}, x)=\Delta_0\psi(\bbox{p}_{\text{F}},
x)$. Here $\psi(\bbox{p}_{\text{F}}, x)$ is a dimensionless normalized
function of the momentum direction and the distance from the surface
($ |\psi(\bbox{p}_{\text{F}}, x)|\le 1$), while $\Delta_0$ is the
maximum value of the bulk order parameter depending upon temperature
and impurity concentration. The spatial dependence of
$\psi(\bbox{p}_{\text{F}},x)$, reduces to a dependence upon
dimensionless coordinate $X=\Delta_0 x/v_F$. Introducing the
dimensionless variables $X,X'$ into Eq.~(\ref{integequat}), we see
that the propagator has the form $g(\bbox{p}_{\text{F}},0)=-i
\sqrt{\tau\Delta_0} G(\bbox{p}_{\text{F}},0)$, where
$G(\bbox{p}_{\text{F}},0)$ does not contain $\tau$ and
$\Delta_0$. Hence, the height of the peak in the density of states is
proportional to $\sqrt{\tau\Delta_0}$. By considering a generalization
of the integral equation Eq.~(\ref{integequat}) to nonzero (although
sufficiently small) values of the energy, one can show analytically
that the shift of the pole position from its zero value is
proportional to $\sqrt{\Delta_0/\tau}$. Thus, introducing the
dimensionless quantity $\Omega_n=\varepsilon_n\sqrt{\tau/\Delta_0}$,
we can represent the contribution of the midgap states to the Green's
function as $g(x=0,\bbox{p}_{\text{F}};\varepsilon_n)=-i
\sqrt{\tau\Delta_0}G(\bbox{p}_{\text{F}},\Omega_n)$.

The relative strength of the influence of impurities on bound states
in the Born and in the unitarity limits can be understood
qualitatively by looking at Eqs.~(\ref{eq:selfenergy-Born}) and
(\ref{eq:Selbstenergie-Unitary}). In the absence of bound states, the
Green's function $g(\bbox{R},\bbox{p}_{\text{F}};\varepsilon_n)$ is
usually quite small (compared to the normal-state value) for
sufficiently small $\varepsilon_n$. In this case, according to
Eqs.~(\ref{eq:selfenergy-Born}) and (\ref{eq:Selbstenergie-Unitary}),
the self-energy function for unitary scatterers can be significantly
greater than the one in the Born limit. By contrast, if there are
bound states on (or quite close to) the Fermi surface, then the
corresponding large pole-like term in the expression for
$g(\bbox{R},\bbox{p}_{\text{F}};\varepsilon_n)$ essentially rises with
decreasing temperature. This leads to the inverse situation, that is
to small values of the self-energy for unitary scatterers as compared
to the Born limit for the same values of the scattering rates. An
analogous conclusion can be drawn for the retarded propagator and the
self-energy function taken at energies close to some bound state, even
if it is not at the Fermi surface. Since the pole-like term decreases
with increasing $\Gamma$, the above consideration, in general, does
not work for sufficiently large values of $\Gamma_b$, $\Gamma_u$. Our
numerical calculations justify the above conclusion: in the presence
of unitary scatterers, for sufficiently small values of $\Gamma_u$,
the bound states are broadened much more weakly than in the Born
limit. In contrast to the Born limit, the dependence of the propagator
on the parameter $\Delta_0/\Gamma_u$ does not reduce to a power-law
behavior in the unitarity limit, so that simple scaling estimates are
not fruitful in this limit. However, some rough qualitative estimates
can be done in the unitarity limit as well, in particular, by
considering the simplest model $d$-wave order parameter, whose
momentum direction dependence reduces to $\Delta_0$ (for
$0<\phi<\pi/2$ and $\pi<\phi<3\pi/2$) and to $-\Delta_0$ (for
$\pi/2<\phi<\pi$ and $3\pi/2<\phi<2\pi$). Then for small enough
$\Gamma_u$ we find the propagator
$g(x=0,\bbox{p}_{\text{F}};\varepsilon_n=0)$ to be proportional to
$\sqrt{\Gamma_u/\Delta_0}\exp(A\Delta_0/\Gamma_u)$ with some numerical
factor $A$ of the order of unity.

We will now concentrate on the temperature and impurity dependences of
the Josephson critical current. Two typical and experimentally
relevant geometries will be studied: the ``symmetric'' junction for
which $\alpha_L=\alpha_R$, and the ``mirror'' junction for which
$\alpha_L=-\alpha_R$.

\begin{figure}[tbp]
\begin{center}
\leavevmode
\psfig{figure=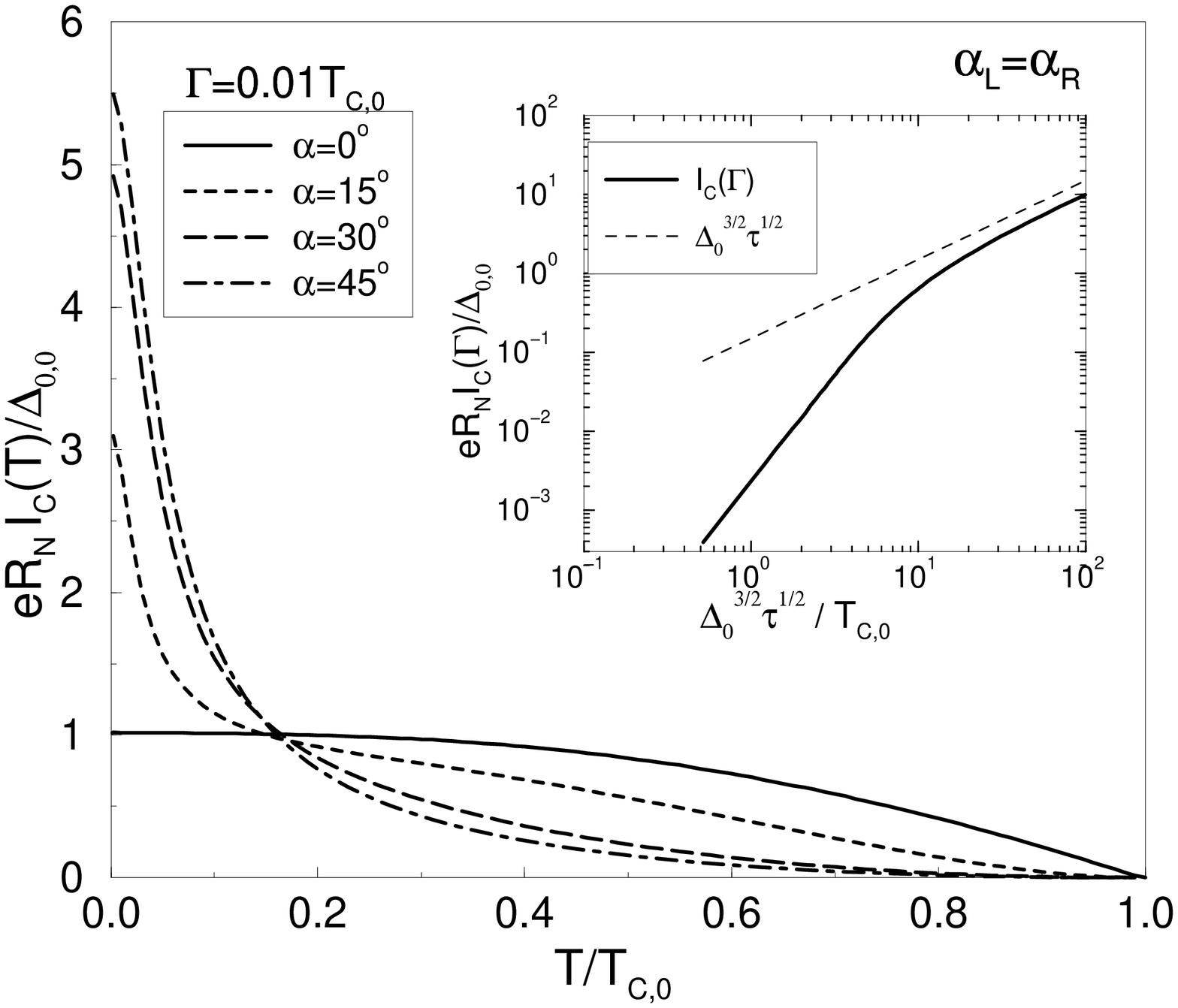,width=7cm,angle=-90}
\psfig{figure=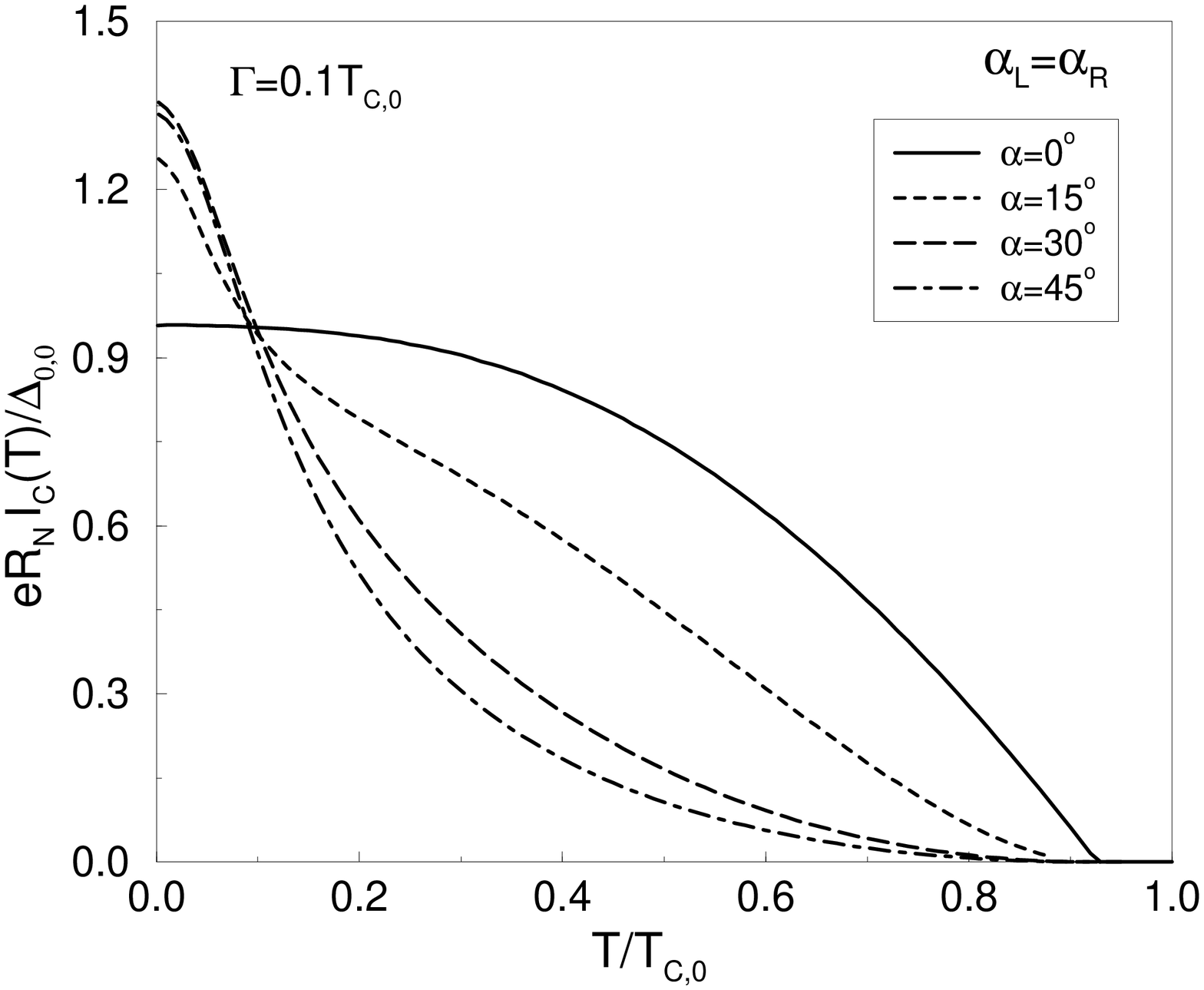,width=7cm,angle=-90}
\narrowtext
\caption{Temperature dependence of the critical current
for the symmetric junction, i.e., $\alpha_L=\alpha_R$ and different
misorientation angles $\alpha$. Impurity scattering is parameterized by the
scattering rate $\Gamma_b$ using the Born approximation,
Eq.~(\protect\ref{eq:selfenergy-Born}).
Inset: Critical current for misorientation $\alpha=45^\circ$ and fixed
temperature $T=0.005T_{C,0}$ as a function of the scattering rate.
}
\label{fig:symborn}
\end{center}
\end{figure}

In contrast to surface roughness, bulk impurities not only broaden the
bound states but also change the maximum of the bulk pair potential
and the critical temperature from their clean-system values
$\Delta_{0,0}$ and $T_{C,0}$. This can be seen in the lower panel of
Fig.~\ref{fig:symborn}, where the finite scattering rate leads to a
renormalized value of $T_C$. Figure~\ref{fig:symborn} shows the
critical current for the symmetric geometry in Born approximation. The
anomalous temperature dependence discussed in
Ref.~\onlinecite{barash96} is still visible, but bulk impurity cuts
off the divergence at zero temperature.

We obtain an analytical estimate of the Josephson critical current in
the zero-temperature limit from Eq.(\ref{eq:critical-current}), taking
into account the large low-energy values of the quantities $f^{L(R)}$,
which are associated with $g$ the same way as in the clean
limit\cite{barash96,barash972}:
$f(x=0,\bbox{p}_{\text{F}};\varepsilon_n)=
f^+(x=0,\bbox{p}_{\text{F}};\varepsilon_n)=
-i\text{sign}(v_x\Delta_{\infty}
\left(\bbox{p}_{\text{F}})\right)g(x=0,\bbox{p}_{\text{F}};\varepsilon_n)$.
Proceeding analogously to Ref.\ \onlinecite{barash96} for the
symmetric (mirror) junction and introducing $\Omega$ as a new
integration variable, we get
\begin{eqnarray}
&&j_C=\pm\frac{e N_0^L}{2\pi^2}\Delta_0^{3/2}\tau^{1/2}\times\nonumber\\
&&\int\limits_0^\infty d\Omega\left\langle D(\bbox{p}_{\text{F} in}^L)
v_{x}^L(\bbox{p}_{\text{F} in}^L)G^2(X=0,\bbox{p}_{\text{F} in}^L,\Omega)
\right\rangle_{\bbox{p}_{\text{F} in}^L}\; .
\label{eq:critical-current2}
\end{eqnarray}
The plus (minus) sign corresponds here to the symmetric (mirror)
junction.

Thus, in the Born limit and under the condition $\Delta_0\tau\gg 1$ the critical
current turns out to be proportional to $\Delta_0^{3/2}\tau^{1/2}$.

\begin{figure}[tbp]
\begin{center}
\leavevmode
\psfig{figure=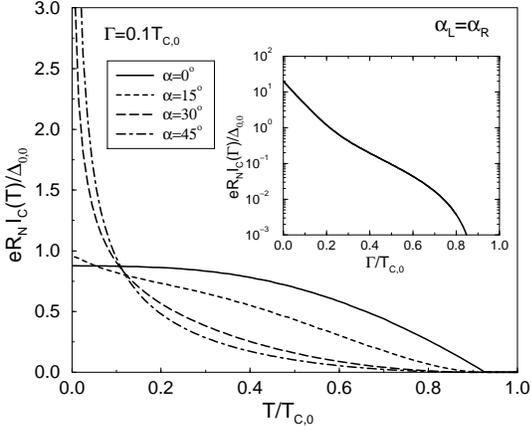,width=7cm,angle=-90}
\narrowtext
\caption{Critical current as a function of temperature for the
symmetric junction in the unitary limit.
Inset: Critical current for misorientation
$\alpha=45^\circ$ and fixed temperature $T=0.01T_{C,0}$ as a function
of the scattering rate.}
\label{fig:symuni}
\end{center}
\end{figure}

In the unitarity limit, shown in Fig.~\ref{fig:symuni}, we find that
the bound states are remarkably stable to impurity scattering,
according to our conclusion made above. For sufficiently large values
of $\Gamma_u$ (when $\Gamma_u /T_{C,0}$ is of the order of unity),
however, the influence can be substantial as can be seen in the inset
to Fig.~\ref{fig:symuni}.

\begin{figure}[htbp]
\begin{center}
\leavevmode
\psfig{figure=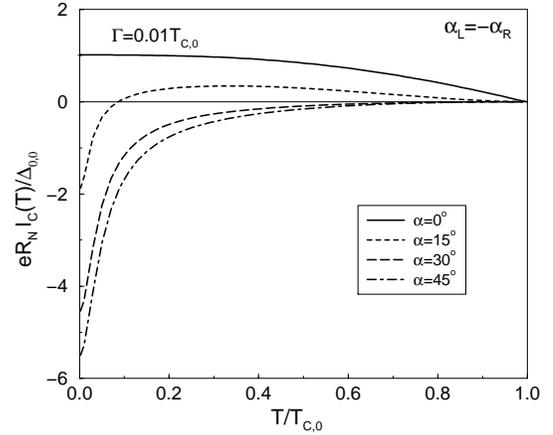,width=7cm,angle=-90}
\psfig{figure=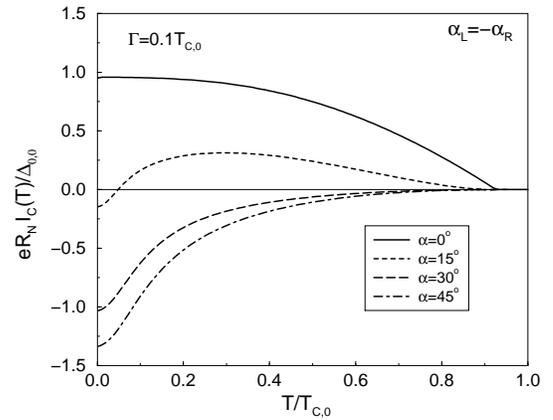,width=7cm,angle=-90}
\narrowtext
\caption{Temperature dependence of the critical current for the mirror
junction, i.e., $\alpha_L=-\alpha_R$ for two scattering rates
$\Gamma_b$ in Born approximation.}
\label{fig:mirborn}
\end{center}
\end{figure}

\begin{figure}[htbp]
\begin{center}
\leavevmode
\psfig{figure=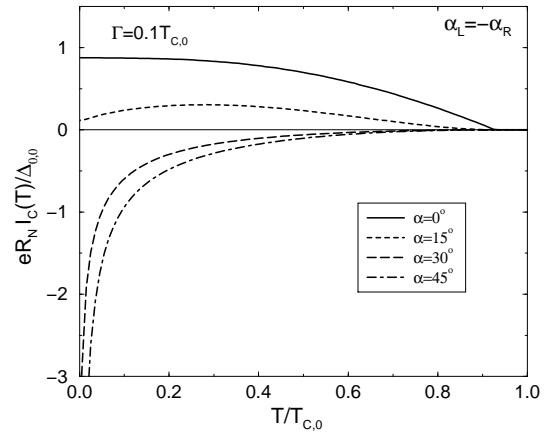,width=7cm,angle=-90}
\narrowtext
\caption{Critical current as a function of temperature for the mirror junction
and scattering in the unitary limit.}
\label{fig:miruni}
\end{center}
\end{figure}

Figures~\ref{fig:mirborn} and \ref{fig:miruni} show the corresponding
results for the mirror junction. As in Ref.~\onlinecite{barash96}, the
critical current changes sign, i.e., the junction changes character
(for some misorientation angles) and becomes a $\pi$-junction at low
temperatures. Bulk impurity scattering weakens this tendency.

One simple model for a surface (or an interface) with roughness is a
thin dirty layer containing Born impurities near the surface (or
around the interface)
\cite{ovchin,culetto,kurki,barash96,rainersauls}. We note that the
dirty layer with unitary scatterers would influence the bound states
and the Josephson critical current less strongly than in the Born
limit (for the same value of the scattering rate). The problem of the
most suitable model for surface and interface roughness for a given
experimental situation is still open in this context.

In conclusion, quasiparticle scattering by bulk impurities as well as
surface roughness results in the broadening of surface (interface)
quasiparticle bound states in tunnel junctions of $d$-wave
superconductors. In this paper, we have studied this broadening due to
bulk impurities and shown that scatterers can essentially reduce the
height of the peak in the density of states and the low-temperature
anomaly in the Josephson critical current. We have shown that bound
states are more sensitive to Born scatterers than to unitary
scatterers at a given value of the scattering rate. Thus, unitary
scatterers would be less detrimental than Born scatterers to the
observability of the low-temperature anomalies in the Josephson
critical current caused by bound states.

We would like to acknowledge discussions with W. Belzig.
Yu.B. and V.I. acknowledge support by the Russian Foundation
for Basic Research under grant No.~96-02-16249.

\end{document}